\documentclass[aps,prc,twocolumn,floatfix,showpacs]{revtex4}

\usepackage{float}
\usepackage{graphicx,amsfonts,amssymb,amsbsy}
\usepackage{amsmath,amsthm,latexsym}
\usepackage{slashed}
\usepackage{ulem}
\usepackage[export]{adjustbox}
\usepackage[colorlinks]{hyperref}
\usepackage{enumitem} 
\usepackage{orcidlink}
\newcommand{\mm}{\mathrm{MM}}

\newcommand{\sm}{\mathrm{SM}}

\newcommand{\sat}{\mathrm{sat}}
\newcommand{\sym}{\mathrm{sym}}

\begin{document}

\title{Gravitational wave asteroseismology of neutron stars with unified EOS: on the role of high-order nuclear empirical parameters}

\author{Guilherme Grams \orcidlink{0000-0002-8635-383X}}
\email{guilherme.grams@ulb.be}
\affiliation{Institut d’Astronomie et d’Astrophysique, Université Libre de Bruxelles, Brussels, Belgium}

\author{César V. Flores \orcidlink{0000-0003-1298-9920}}
\email{cesar.vasquez@uemasul.edu.br}
\affiliation{Universidade Estadual da Região Tocantina do Maranhão, UEMASUL, Centro de Ciências Exatas, Naturais e Tecnológicas, Imperatriz, CEP 65901-480 Maranhão, Brazil}
\affiliation{Universidade Federal do Maranhão, UFMA, Departamento de Física-CCET, Campus Universitário do Bacanga, São Luís, CEP 65080-805 Maranhão, Brazil}

\author{César H. Lenzi \orcidlink{0000-0001-5887-338X
}}
\email{chlenzi@ita.br}

\affiliation{
\mbox{$^2$Departamento de F\'isica, Instituto Tecnol\'ogico de Aeron\'autica, DCTA, 12228-900, 
S\~ao Jos\'e dos Campos, SP, Brazil} \\ 
}

\begin{abstract}
We analyze the sensitivity of non-radial fluid oscillation modes and tidal deformations in neutron stars to high-order nuclear empirical parameters (NEP). In particular, we study the impact of the curvature and skewness of the symmetry energy $K_{\rm sym}$, $Q_{\rm sym}$, and the skewness of the binding
energy in symmetric nuclear matter $Q_{\rm sat}$. As we are interested in the possibility of gravitational wave detection by future interferometers, we consider that the tidal interaction is the driving force for the quadrupolar non-radial fluid oscillations. We have also studied the correlations between those quantities, which will be useful to understand the strong physics of gravitational wave phenomena. Our main results show that $K_{\rm sym}$ impacts the frequencies of the fundamental mode mainly for low-mass neutron stars. The NEP $Q_{\rm sym}$ and $Q_{\rm sat}$ affect the fundamental modes of intermediate and heavy neutron stars, respectively. In the case of the first pressure mode, $K_{\rm sym}$ shows a small effect, while $Q_{\rm sat}$ shows a considerable decrease in this oscillation mode independent of the neutron star mass. Similarly, for tidal deformability, the NEP $Q_{\rm sat}$ and $Q_{\rm sym}$ show a bigger impact than $K_{\rm sym}$. Given the impact of the NEP on gravitational wave phenomena and the currently large uncertainties of these parameters, the prospect of higher sensitivity in future gravitational wave detectors promise a possible new tool to constrain high-order NEP.
\end{abstract}


\maketitle

\section{Introduction}

The dawn of the multimessenger era has begun with the binary neutron star (NS) merger observation of gravitational waves (GW)~\cite{AbbottLigo_2017} and its electromagnetic counterpart~\cite{Cowperthwaite_2017}, and it is providing new insights into the nature of matter at extreme conditions.
This new and exciting field of research is leading to huge investments all around the globe in the planning of third-generation (3G) interferometers, such as the Einstein Telescope~\cite{Punturo_2010,Branchesi_2023} and the Cosmic Explorer~\cite{Hall21,Hall22}, which will observe the Universe by GW at unprecedented distances. The precise modeling of NS and their mergers is a prerequisite to tackling fundamental questions that will be studied by 3G detectors, e.g., the origin of the Universe, NS oscillations, rotational oscillations modes of compact star and the equation of state (EoS) of dense matter. 

The dense matter EoS is a key ingredient to model NS physics.
Describing all the different NS regions within a unified framework is a challenge in view of the very wide range of densities spanned in its different layers. 
For example at the surface of the star, we can find Fe$^{56}$ and neighboring elements, on the outer crust the atomic nuclei embedded on a gas of electrons becomes heavier and rich in neutrons until some neutrons start to drip from the nuclei. This marks the beginning of the NS inner crust, where exotic neutron rich nuclei coexist with a neutron (super) fluid~\cite{Chamel08b}. 
At even more high densities we reach the outer core which is composed of free leptons and nucleons. 
The composition of the NS inner core is currently unknown, and the modeling of this region can maintain the nucleonic hypothesis or add extra degrees of freedom like hyperons, meson condensates, and quark matter~\cite{Oertel17,Blaschke2018,Somasundaram23}.

Due to the difficulty of constructing an EoS that describes all the NS within an unified approach, specially due to the complexity of the crust physics, a common practice is to match different models for the NS crust and core. 
It was however shown that the use of non-unified EoS adds extra uncertainties on NS observables, e.g $4\%$ on radius, and $30\%$ crust thickness~\cite{Fortin16}. 
For GW phenomena, Ref.~\cite{Suleiman21} shows that an inconsistent treatment of crust and core can lead to $20\%$ error on NS tidal deformability with respect to the unified treatment and the authors of~\cite{Ferreira20} found that different  crust-core glueing leads to $20\%$ difference on the Love number $k_2$ when compared to unified EoS.
To avoid adding uncertainties related to different crust-core matching, in the present work we use thermodynamic consistent and unified EoS, on the lines of Ref.~\cite{Grams22a}.

Several works have analyzed the possibility of extra degrees of freedom on the NS interior and its effect on GW seismology, such as the impact of hyperons onset~\cite{Pradhan21}, phase transition to quark matter~\cite{Kumar23} and the effect of dark matter admixed NS~\cite{Shirke24}.
However more exotic particles might be present on compact stars, current observations do not rule out the possibility of NS composed only by nuclear matter~\cite{Dinh21,Somasundaram23,Mondal23} due to the masquerade effect~\cite{Alford05}.
In the present work we keep the nucleonic hypothesis to better understand the impact on high-order nuclear empirical parameters (NEP) on NS.

As we have seen in the previous discussion, a NS can have different composition in its interior, 
and considerable effort is being made to better understand the complexities of the NS EoS.
In this respect, the study of the fundamental mode ($f$-mode) has attracted the attention by the physics community, because it can bring important information about the mean density of a NS. The $f$-mode belongs to the family of non radial oscillations modes of a NS, and this mode is related to the mean density of the star. Another interesting oscillation mode is the $p$-mode, which has higher frequencies than the $f$-mode. The $f$-mode has received much attention because it can be excited more easily, however recent simulations of supernovas  show that the $p$-mode could also be activated \cite{10.1093/mnras/stae834} by the strong dynamics of the explosion. We also have the $g$-mode, which is related to density discontinuities inside the star, and finally, when rotation is considered, it can appear the $r$-mode. All of these modes are very important because are coupled to the emission of gravitational radiation, and it is very clear to the  scientific community that the detection of those modes will be imminent when the 3G detectors will be finished and operating. Another phenomena that can be observed in a GW signal is the tidal deformability of a NS. Such deformations manifest when two NS are very close and can be coupled to non radial oscillations. The NS tidal deformability is very important because is directly related to the matter composition of NS. During the late inspiral, tidal deformations raised on each star by the gravitational field of its companion depend crucially on the star’s internal properties \cite{PhysRevD.99.083014,PhysRevD.81.123016}.

Recently there have been different works trying to correlate $f$ and $p$-modes, tidal deformability, mass and star radius. For example we can mention \cite{Kunjipurayil22}, where it was used  Skyrme Hartree-Fock and relativistic mean-field models. Their study shows a strong correlation between the frequencies of $f$- and $p$-modes, damping times with the pressure of $\beta$-equilibrated matter. Also in \cite{Montefusco24}, the authors considered a metamodel representation of the nucleonic EoS of an NS, with the objective to identify possible biases on oscillation modes arising from asuming a barotropic EoS, and provided a
distribution of expected $f$ and $p$ mode frequencies that could be detected by 3G
GW interferometers.

 In this work we study the effect of the high-order NEP of the EoS on the non radial oscillations and tidal deformations. For this purpose, we construct a set of unified (crust-core) EoS with a flexible approach thanks to the meta-modeling technique~\cite{Margueron2018a}. Such technique was already largely used to investigate uncertainties on NS mass and radii~\cite{Margueron2018b}. For example, the authors of Ref.~\cite{Sukrit_2021,Pradhan_2023} investigate the role of nuclear parameters on NS $f$-modes, however, restricted to lower order parameters, such as the effective mass and the nuclear symmetry energy.

 The paper is organized as follows. In section~\ref{sec:EOS} we present the method to compute the EoS used in this work. In section~\ref{sec:nonradosci} we present the standard theoretical basis of non radial oscillations together with our results for the $f$- and $p$-modes. In section ~\ref{sec:tidal} we make a discussion about tidal deformations and finally in section ~\ref{sec:conclusion} we make our conclusions.

\section{Equations of state} 
\label{sec:EOS}

We use unified EoS in which the crust is consistently computed alongside the core EoS using a compressible liquid drop model (CLDM) approach as detailed in Ref.~\cite{Grams22a,Grams22b}. We choose as reference EoS  the H2$_{MM}$ which is the meta-model version of the chiral Hamiltonian H2 proposed in Ref.~\cite{Drischler2016}. We choose H2 among the seven Hamiltonians proposed in Ref.~\cite{Drischler2016} since it is the Hamiltonian that show best agreement with experimental nuclear masses in the recent study~\cite{Grams22a,Grams:2021a}.  

The topological properties of the nuclear matter energy per particle can be written as a Taylor expansion around saturation density, 
\begin{eqnarray}
e_\sm(n) &=& E_\sat + \frac 1 2 K_\sat x^2 + \frac{1}{6} Q_\sat x^3 
+ \frac{1}{24} Z_\sat x^4+\dots \, , \nonumber \\ \\
e_\sym(n) &=& E_\sym + L_\sym x + \frac 1 2 K_\sym x^2 + \frac{1}{6} Q_\sym x^3 \nonumber \\ 
&&\hspace{1cm}+ \frac{1}{24} Z_\sym x^4+\dots \, 
\end{eqnarray}
for the binding energy per particle in symmetric matter and the symmetry energy respectively. The density expansion parameter is defined as $x=(n-n_\sat)/(3n_\sat)$ and $E_\sat$, $K_\sat$, $E_\sym$, $L_\sym$, etc, are the NEP.

Inspired on the above Taylor expansion, the nuclear meta-model (MM)~\cite{Margueron2018a} was created to connect the energy density functional (EDF) with nuclear experiments and ab-initio theory within a flexible approach. The MM is an EDF that does not make assumptions on the underlying nuclear force, similar to agnostic models (polytropic, sound speed). It, however, allows predictions of matter composition, therefore it belongs to the semi-agnostic class of functionals. 
The flexible characteristic of the MM makes it suitable for mimicking existing functionals, as it was done for example in~\cite{Dinh23,Grams22b}.
In Ref.~\cite{Grams22b} the MM was fitted to chiral Effective Field Theory (EFT) Hamiltonians from~\cite{Drischler2016}. Since chiral EFT data is valid up to $\approx 1-2 n_\sat$, only the low-order NEP are constrained by chiral EFT data.
At Ref.~\cite{Grams22b}, chiral EFT data was used to constrain $E_{\sat / \sym}$ up to $K_{\sat / \sym}$, while the high order ones was chosen arbitrarily with the only requirement of having an EoS stiff enough to sustain 2$M_\odot$ pulsars.

\begin{table*}
\caption{\label{tab:NEPH2}Nuclear empirical parameters from the chiral Hamiltonian H2 and used as standard model in the present work. The lowest order NEP are defined at $n_\sat$ ($E$, $L$/$n$, $K$), while the higher order ones ($Q$, $Z$) were fixed to describe 2~M$_\odot$ NS with this model. The last two columns show the effective mass $m^*_\sat$ at saturation in symmetric matter and the effective mass splitting $\Delta m^*_\sat$, see Ref.~\cite{Grams22a,Grams22b} for more details.}
\begin{ruledtabular}
\begin{tabular}{ccccccccccccc}
Model & $E_\sat$ & $n_\sat$ & $K_\sat$ & $Q_\sat$& $Z_\sat$  & $E_\sym$ &  $L_\sym$ & $K_\sym$ & $Q_\sym$ & $Z_\sym$ & $m^*_\sat$ & $\Delta m^*_\sat$ \\
 &(MeV) & (fm$^{-3}$) &(MeV) & (MeV) & (MeV) & (MeV) & (MeV) & (MeV) & (MeV) & (MeV) & ($m_N$) & ($m_N$) \\ \hline
H2$_\mm$ &  -15.8 & 0.176 & 237 & -220 & -200 & 32.0 & 43.9 & -144 & 700 & 500  & 0.61 & 0.41 \\
\end{tabular}
\end{ruledtabular}
\end{table*}

At the present work we take advantage of the flexibility of the MM and vary the high-order parameters to investigate their role in GW asteroseismology of NS. In particular, we investigate the impact of variations on the isoscalar parameter $Q_{\rm sat}$ and isovector $K_{\rm sym}$ and $Q_{\rm sym}$. The NEP of H2$_{MM}$ are shown in Table~\ref{tab:NEPH2}. Note that the unique flexible characteristic of the MM allow us to vary independently each NEP while keeping all other parameters fixed to the original model, which is not possible to traditional EDF such as Skyrme, Gogny or relativistic mean-field models.

The NEP $K_{\rm sym}$ has currently a large uncertainty and different studies have lead to negative values $\approx - 100$ MeV ($\pm 100$ MeV)~\cite{Tews2017,Sagawa2019,Grams22c}. We therefore vary $K_{\rm sym}$ within this range of uncertainty, and created five new EoS with $K_{\rm sym} = -200, -150, -100, -50,$ and -1 MeV for the present work.

Experimental data and ab-initio predictions are even more scarce for $Q_{\rm sat}$ and $Q_{\rm sym}$. These NEP are currently unconstrained, and we take guidance by Ref.~\cite{Margueron2018a} to vary their values around the reference model H2$_{MM}$ (Tab.~\ref{tab:NEPH2}). 
To investigate the impact of these uncertainties on GW asteroseismology, we create new EoS with $Q_{\rm sat} = -400, 0, 220,$ and $400$ MeV, and $Q_{\rm sym} = 100, 400,$ and $1000$ MeV.
Note that the reference EoS H2$_{MM}$ has $K_{\rm sym} =-144$ MeV, $Q_{\rm sat}=-220$ MeV and $Q_{\rm sym}=700$ MeV.

\begin{figure}[h!]
 \centering
 \includegraphics[angle=0,scale=0.57]{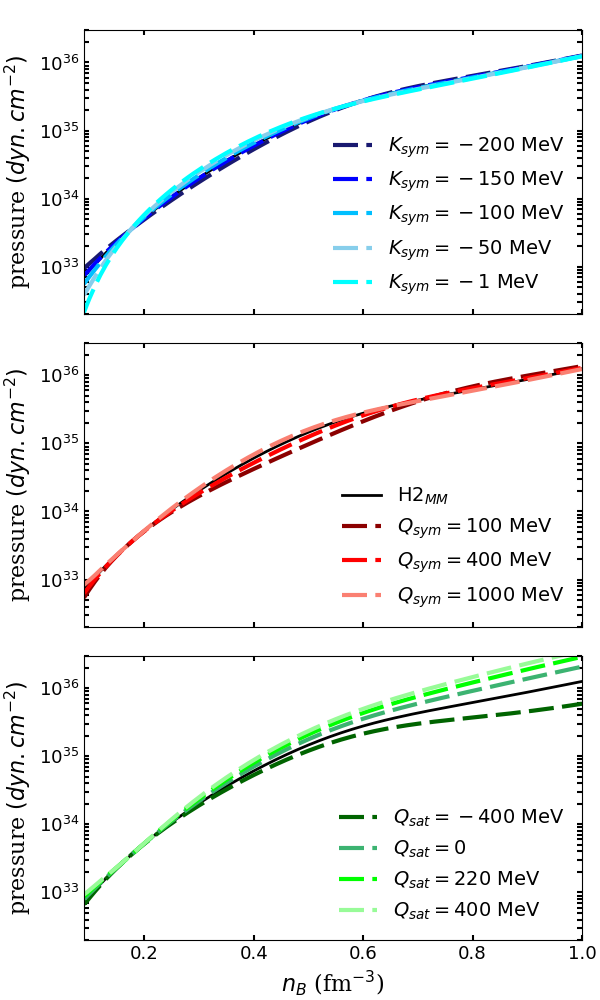} 
\caption{Equation of state for the models used in the present work. The top panel shows the effect of varying $K_\sym$ individually, while the effect of varying $Q_\sym$ is displayed in the center and of varying $Q_\sat$ in the bottom panel. The reference EoS H2$_{MM}$ is shown in all panels with solid line.}
\label{fig:eos}
\end{figure}

We show in Fig.~\ref{fig:eos} the EoS, i.e., pressure with respect to baryon density, for all models used in the present work. Note that each NEP varied independently impacts the stiffness of the EoS in different density regions: top panel shows that $K_{\rm sym}$ has a small impact at densities $\approx 0.3$ fm$^{-3}$, while middle panel shows an impact of $Q_{\rm sym}$ at 
densities $\approx 0.5$ fm$^{-3}$. Bottom panel shows that $Q_{\rm sat}$ impacts the EoS at densities higher than 0.6 fm$^{-3}$. For all NEP we note an stiffer EoS for more positive their values.

In summary, we start with the base H2$_{MM}$ EoS constructed in Ref.~\cite{Grams22a}  and created twelve new EoS where the isoscalar $Q_{\rm sat}$ and isovectors $K_{\rm sym}$ and $Q_{\rm sym}$ NEP were varied independently.
All EoS constructed for the present work are unified (crust-core) and assume neutron-proton-electron-lepton matter in $\beta$-equilibrium on the NS interior.

\section{non-radial oscillations of a compact star}
\label{sec:nonradosci}

\subsection{The equilibrium configuration}

The equilibrium configuration of a non-rotating, perfect fluid compact star is governed by the Tolman-Oppenheimer-Volkoff (TOV) equations, which describe the star's relativistic structure. To derive these equations, we begin by selecting a spherically symmetric static metric, as follows:
\begin{equation}
    ds^2 = - e^{\nu}dt^2 + e^{\lambda} dr^2 + r^2(d\theta^2 + sin^2 \theta d \phi^2),
\end{equation}
and for matter we consider the energy-momentum tensor of a perfect fluid
\begin{equation}\label{tem}
    T_{\mu\nu}=p g_{\mu\nu}+(\epsilon + p)u_{\mu}u_{\nu},
\end{equation}
where $\rho$, $p$, and $u_{\mu}$ are respectively the energy density, pressure, and the fluid's four-velocity of a fluid element. 

The two equations above are used as input in the well know Einstein field equations
\begin{equation}
 G_{\mu\nu}=8\pi T_{\mu\nu}\label{EFeq},
\end{equation}
and after some algebraic manipulation, it is possible to obtain the following relativistic structure equations 
\begin{eqnarray}
\label{tov1}
p' &=& - \frac{\epsilon m}{r^2}\bigg(1 + \frac{p}{\epsilon}\bigg)
	\bigg(1 + \frac{4\pi p r^3}{m}\bigg)\left(1 - \frac{2m}{r}\right)^{-1},  \\
\label{tov2}
\nu' &=& - \frac{2}{\epsilon} p^{'} \bigg(1 + \frac{p}{\epsilon}\bigg)^{-1}, \\
\label{tov3}
m'& =& 4 \pi r^2 \epsilon,
\end{eqnarray}
where $m$ is the gravitational mass inside the radius $r$  and $\nu(r)$ is the metric potential related to the local gravitational field. The pressure $p$ and the energy density $\epsilon$ are related by the EoS given before. The integration of the the structure equations use the initial conditions: $m(0) =0$, $p(R) =0$, and the mass $M$ and radius $R$ of the star is obtained when the integration stops at the surface of the star, where the pressure vanishes.

Also we would like to comment that for matching of quantities inside and outside the star we use the additional condition:  $\nu(R)= \ln ( 1- {2M}/{R} )$. This condition is very important in the numerical work with the non radial oscillations equations (which are very important in GW asteroseismology) and in the treatment of the equations of tidal deformability.

\subsection{The oscillation equations}

The polar non-radial perturbations corresponding to the oscillations of a non-rotating perfect fluid star can be described through a set of equations presented in \cite{1983ApJS...53...73L, 1985ApJ...292...12D}.  In this formalism the perturbed metric tensor  reads
\begin{eqnarray}
ds^2 & = & -e^{\nu}(1+r^{\ell}H_0Y^{\ell}_{m}e^{i\omega t})dt^2  - 2i\omega r^{\ell+1}H_1Y^{\ell}_me^{i\omega t}dtdr   \nonumber \\
&& + e^{\lambda}(1 - r^{\ell}H_0Y^{\ell}_{m}e^{i \omega t})dr^2 \nonumber \\
&& + r^2(1 - r^{\ell}KY^{\ell}_{m}e^{i \omega t})(d\theta^2 + \sin^2\theta d\phi^2),
\end{eqnarray}
and the polar perturbations in the position of the fluid elements are given  by the following Lagrangian displacements
\begin{eqnarray}
\xi^{r} &=& r^{\ell-1}e^{-\lambda/2}WY^{\ell}_{m}e^{i\omega t}, \\
\xi^{\theta} &=& -r^{\ell - 2}V\partial_{\theta}Y^{\ell}_{m}e^{i\omega t}, \\
\xi^{\phi} &=& -r^{\ell}(r \sin \theta)^{-2}V\partial_{\phi}Y^{\ell}_{m}e^{i\omega t} ,
\end{eqnarray}
where $Y^{\ell}_{m}(\theta,\phi)$ are the spherical harmonics.

Using the previous quantities  and putting them inside the perturbed Einstein equations, we can obtain the following set of first order linear differential equations that govern the oscillations of the star \cite{1985ApJ...292...12D}:
\begin{eqnarray}
H_1' &=&  -r^{-1} [ \ell+1+ 2Me^{\lambda}/r +4\pi   r^2 e^{\lambda}(p-\epsilon)] H_{1}   \nonumber \\ 
&& +  e^{\lambda}r^{-1}  \left[ H_0 + K - 16\pi(\epsilon+p)V \right] ,      \label{osc_eq_1}  \\
 K' &=&    r^{-1} H_0 + \frac{\ell(\ell+1)}{2r} H_1   - \left[ \frac{(\ell+1)}{r}  - \frac{\nu'}{2} \right] K  \nonumber \\
&&  - 8\pi(\epsilon+p) e^{\lambda/2}r^{-1} W \:,  \label{osc_eq_2} \\
 W' &=&  - (\ell+1)r^{-1} W   + r e^{\lambda/2} [ e^{-\nu/2} \gamma^{-1}p^{-1} X    \nonumber \\
&& - \ell(\ell+1)r^{-2} V + \tfrac{1}{2}H_0 + K ] \:,  \label{osc_eq_3}
\end{eqnarray}

\begin{eqnarray}
X' &=&  - \ell r^{-1} X + \frac{(\epsilon+p)e^{\nu/2}}{2}  \Bigg[ \left( r^{-1}+{\nu'}/{2} \right)H_{0}  \nonumber  \\
&& + \left(r\omega^2e^{-\nu} + \frac{\ell(\ell+1)}{2r} \right) H_1    + \left(\tfrac{3}{2}\nu' - r^{-1}\right) K   \nonumber  \\
&&    - \ell(\ell+1)r^{-2}\nu' V  -  2 r^{-1}   \Biggl( 4\pi(\epsilon+p)e^{\lambda/2} \nonumber  \\
&& + \omega^2e^{\lambda/2-\nu}  - \frac{r^2}{2}  (e^{-\lambda/2}r^{-2}\nu')' \Biggr) W \Bigg]  \:,
\label{osc_eq_4}
\end{eqnarray}
where the prime denotes a derivative with respect to $r$ and  $\gamma$ is the adiabatic index which that read as
\begin{equation}
    \gamma = \frac{(\epsilon + p)}{p} \bigg( \frac{dp}{d\epsilon} \bigg),
\end{equation}

also, in the equations above the function $X$ is given by
\begin{eqnarray}
X =  \omega^2(\epsilon+p)e^{-\nu/2}V - \frac{p'}{r}e^{(\nu-\lambda)/2}W   \nonumber \\
  + \tfrac{1}{2}(\epsilon+p)e^{\nu/2}H_0 ,
\end{eqnarray}
and $H_{0}$ fulfills the algebraic relation 
\begin{eqnarray}
a_1  H_{0}= a_2  X -  a_3 H_{1}  + a_4 K,
\end{eqnarray}
with
\begin{eqnarray}
a_1 &=&  3M + \tfrac{1}{2}(l+2)(l-1)r + 4\pi r^{3}p  ,  \\
a_2 &=&  8\pi r^{3}e^{-\nu /2}   , \\
a_3 &=&  \tfrac{1}{2}l(l+1)(M+4\pi r^{3}p)-\omega^2 r^{3}e^{-(\lambda+\nu)}  ,  \\
a_4 &=&  \tfrac{1}{2}(l+2)(l-1)r - \omega^{2} r^{3}e^{-\nu}  \nonumber \\
    &&    -r^{-1}e^{\lambda}(M+4\pi r^{3}p)(3M - r + 4\pi r^{3}p)   .
\end{eqnarray}

Outside the star, the perturbations equations on the fluid are null and the differential equations reduce to the very well know Zerilli equations for the metric vacuum perturbations, which can expressed aas follows
\begin{equation}
\frac{d^{2}Z}{dr^{*2}}=[V_{Z}(r^{*})-\omega^{2}]Z,
\end{equation}
where $Z(r^{*})$ and $dZ(r^{*})/dr^{*}$  are related to the metric perturbations  $H_{0}(r)$ and $K(r)$ by the transformations given in Refs. \cite{1983ApJS...53...73L,1985ApJ...292...12D}.  We can also note the ``tortoise'' coordinate given by 
\begin{equation}
r^{*} = r + 2 M \ln (r/ (2M) -1), 
\end{equation}
and  the effective potential  $V_{Z}(r^{*})$ is given by
\begin{eqnarray}
V_{Z}(r^{*}) = \frac{(1-2M/r)}{r^{3}(nr + 3M)^{2}}f(r),
\end{eqnarray}
\begin{equation}
f(r)=[2n^{2}(n+1)r^{3} + 6n^{2}Mr^{2} + 18nM^{2}r + 18M^{3}]
\end{equation}
with $n= (l-1) (l+2) / 2$.

\begin{figure*}[tb]
 \centering
 \includegraphics[angle=0,scale=0.6]{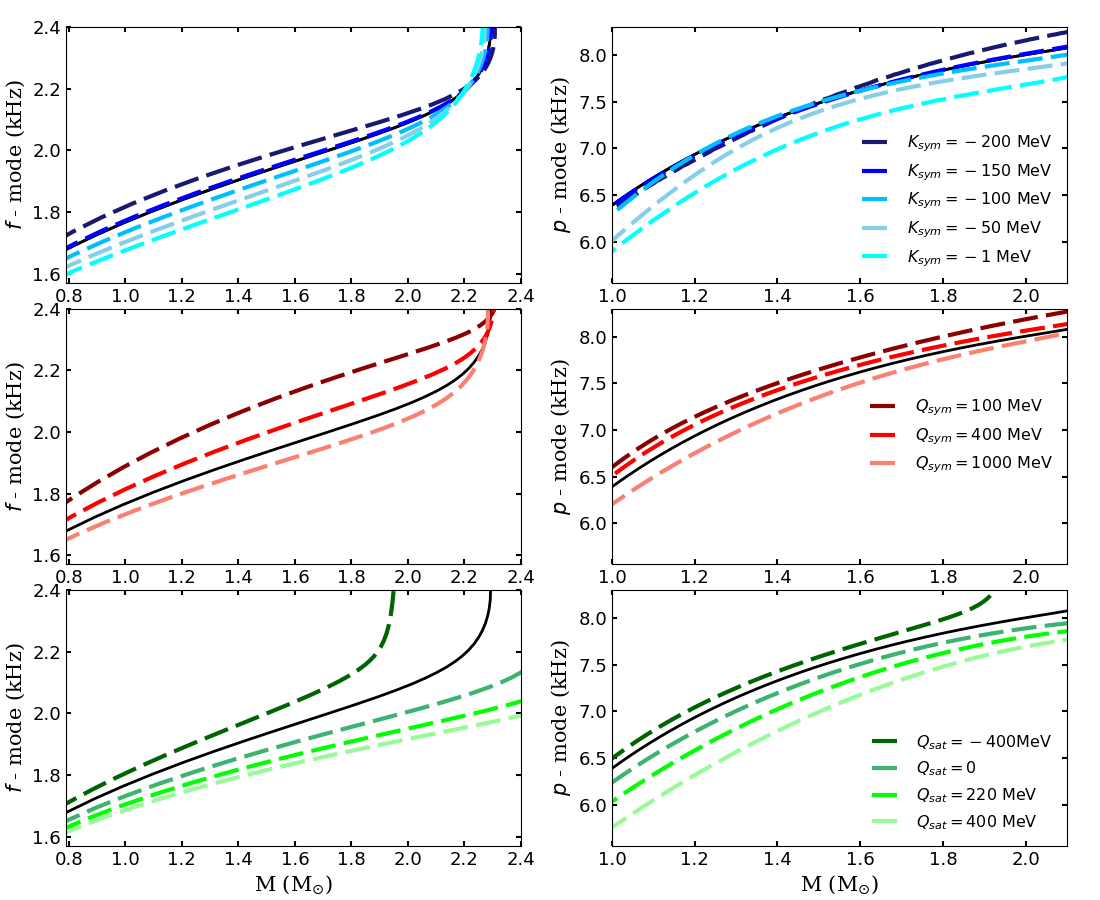} 
\caption{Left panels show the $f$-modes while right panels show the $p$-modes with respect to the NS mass for the EoS used in the present work. The top panel shows the effect of varying $K_\sym$ individually, while the effect of varying $Q_\sym$ is displayed in the center and of varying $Q_\sat$ in the bottom panel. The reference EoS H2$_{MM}$ is shown in all panels with solid line.}
\label{fig:fmodNEP}
\end{figure*}

The system of Eqs. (\ref{osc_eq_1})$-$(\ref{osc_eq_4}) has four linearly independent solutions for given values of $l$ and $\omega$ (we restrict to the $l = 2$ component, which dominates the emission of GW).  

The physical solution needs to verify the following appropriate boundary conditions:

    \begin{enumerate}[label=(\roman*)]
        \item The perturbation functions must be finite everywhere, in particular at $r = 0$. To implement such condition  it is necessary to use a power series expansion of the solution near the singular point $r=0$. The procedure is explained in detail in \cite{1983ApJS...53...73L}.
        
        \item The next boundary condition says that the Lagrangian perturbation in the pressure has to be zero at the surface of the star $r = R$. This implies that the function $X$ must vanish at $r = R$. 
        
        \item And finally, outside the star the perturbed metric describes a combination of outgoing and ingoing GW.  The physical solution of the Zerilli equation is the one that describes purely outgoing gravitational radiation at $r=\infty$.
        
    \end{enumerate}

Such boundary condition cannot be verified by any value of $\omega$ and the frequencies that fulfill this requirement represent the quasinormal modes of the stellar model. The numerical procedure to solve the above  equations is explained in \cite{1983ApJS...53...73L, 1985ApJ...292...12D}.

In Fig.~\ref{fig:fmodNEP} we show the results for the $f$-mode and $p$-mode. In all the figures the vertical axis means the frequency in units of kilohertz and on the horizontal axis we have the stellar mass in units of solar masses. 

On all the left panels we show our results for the fundamental mode. As we can see, in a qualitative form, the increase in the parameter $K_{\rm sym}$ produces a systematic decrease in the frequencies of the fundamental modes. The same effect can be observed for the $Q_{\rm sym}$ and $Q_{\rm sat}$ parameters. We have to note that the effect is the weakest for the case of the $K_{\rm sym}$ parameter and stronger for the case of the $Q_{\rm sat}$ parameter. In all the models we can see a nearly linear behavior of the frequency as a function of the mass, but this happen only just before the models reach the maximum mass. On the other hand, we can observe that the GW frequency is greater than 1.8 kHz for stars with masses above 1.4 M$_{\odot}$. Also, we can see that the maximum frequency is approximately 2.3 kHz.

On all the right panels we have the results for the frequency of the $p$-mode. We note similar effects, i.e., the increase in the parameters tends to decrease the GW frequency. The effect is stronger for the $Q_{\rm sat}$ parameter, however we have to mention that it is difficult to see the effects for the case of the $K_{\rm sym}$ parameter. For all the models the frequency is above approximately 7.0 kHz for stars with masses greater then 1.4 M$_{\odot}$.

\section{Tidal Deformability}
\label{sec:tidal}

We consider in this work that non radial oscillations and tidal deformations exist simultaneously. Then it is possible to obtain very precious information about both phenomena. In recent years, the theory of relativistic tidal effects in binary systems has been the focus of intense research \cite{10.1093/mnras/270.3.611,PhysRevD.96.083005}. Below, we  provide a summary of the procedure for computing the dimensionless tidal parameter $\Lambda$, which measures how easily an object is deformed by an external tidal field. It is defined as follows:
\begin{equation}
 \Lambda = \frac{2k_2}{3C^5} , \label{stidal}
\end{equation}
where $C = GM/R$ is the compactness and $M$ and $R$, are the mass and  radius of the compact object. The parameter $k_2$ is known as the second-order Love number and is given by:
\begin{equation}
\begin{aligned} 
k_{2}= & \frac{8 C^{5}}{5}(1-2 C)^{2}\left[2-y_{R}+2 C\left(y_{R}-1\right)\right] \\ & \times\left\{2 C\left[6-3 y_{R}+3 C\left(5 y_{R}-8\right)\right]\right. \\ & +4 C^{3}\left[13-11 y_{R}+C\left(3 y_{R}-2\right)+2 C^{2}\left(1+y_{R}\right)\right] \\ & \left.+3(1-2 C)^{2}\left[2-y_{R}+2 C\left(y_{R}-1\right)\right] \ln (1-2 C)\right\}^{-1}
\end{aligned}
\end{equation}
where the quantity $y_R=y(r=R)$ is obtained at the surface of the star by means of the integration of the following equation
\begin{equation}
 r\frac{dy}{dr} +y^2 + yF(r) +r^2Q(r) = 0, \label{EL15}
\end{equation}
where the coefficients $F(r)$ and $Q(r)$ are given by:
\begin{eqnarray} 
F(r) &= & {\left[1-4 \pi r^{2}(\epsilon-p)\right]\left[1-\frac{2 m}{r}\right]^{-1} } , \\ 
Q(r) & = & 4 \pi\left[5 \epsilon+9 p+\frac{\epsilon+p}{c_{s}^{2}}-\frac{6}{4 \pi r^{2}}\right]\left[1-\frac{2 m}{r}\right]^{-1}   \nonumber \\ 
& & -\frac{4 m^{2}}{r^{4}}\left[1+\frac{4 \pi r^{3} p}{m}\right]^{2}\left[1-\frac{2 m}{r}\right]^{-2} ,
\label{EL17}
\end{eqnarray}
we also have to mention that $c_{s}^{2} \equiv d p / d \epsilon$ is the squared speed of sound. The boundary condition for Eq. \eqref{EL15}  at $r = 0$ is given by $y(0) = 2$. To obtain the tidal Love number we use the EoS as an input and integrate the TOV equations along with Eq. \eqref{EL15}.

\begin{figure}[h!]
 \centering
 \includegraphics[angle=0,scale=0.55]{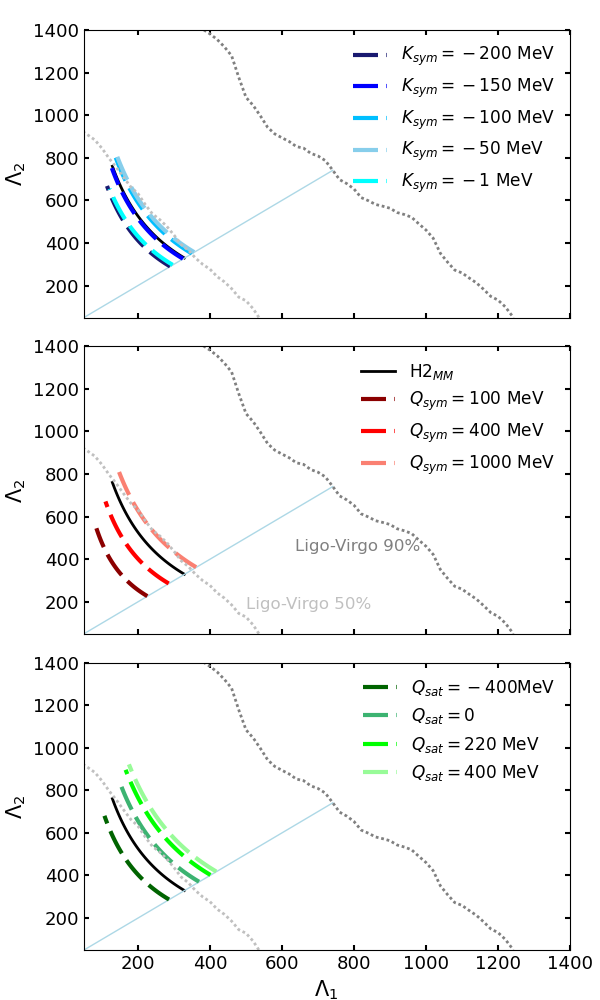} 
\caption{Tidal deformability for the EoS used in the present work. The top panel shows the effect of varying $K_\sym$ individually, while the effect of varying $Q_\sym$ is displayed in the center and of varying $Q_\sat$ in the bottom panel. The reference EoS H2$_{MM}$ is shown in all panels with solid black line. Contours in gray dotted lines show the observational data GW170817~\cite{AbbottLigo_2017}.}
\label{fig:lambdaNEP}
\end{figure}

We show in Figure~\ref{fig:lambdaNEP} the NS tidal deformability computed with the different EoS produced in the present work. The gray dotted lines delineate the 50$\%$ and 90$\%$ confidence regions from the GW170817 event.~\cite{AbbottLigo_2017}. As can be observed, part of the data is below the 50$\%$ confidence line; however, all the EoS presented here are within the current observational uncertainty of LIGO-VIRGO collaboration~\cite{AbbottLigo_2017}. Very stiff EoS are disfavored in this observational scenario~\cite{Lourenco19,Lourenco20}, as greater stiffness results in higher deformability values, which could lead the curves outside the 90$\%$ confidence line. In this sense, it is possible to see the different values for $K_\sym$ shows a small impact, with the most negative values going to lower values of $\Lambda$ and the model with $K_\sym = -1$ MeV getting on top of the observation counter of Ligo Virgo with 50$\%$ of confidence. 

At the middle panels we note that higher values of $Q_\sym$ produce higher values of $\Lambda$, with $Q_\sym = 1000$ MeV being place slightly inside the two observational contours. The bottom panel show the variations of $Q_\sat$, where we note an impact if increasing $\Lambda$ with higher values of $Q_\sat$, where the models with $Q_\sat \geq 0$ are place inside both contour lines, and the negative values produce tidal deformability below the 50$\%$ line of Ligo Virgo.


\begin{figure*}[h!]
 \centering
 \includegraphics[angle=0,scale=0.6]{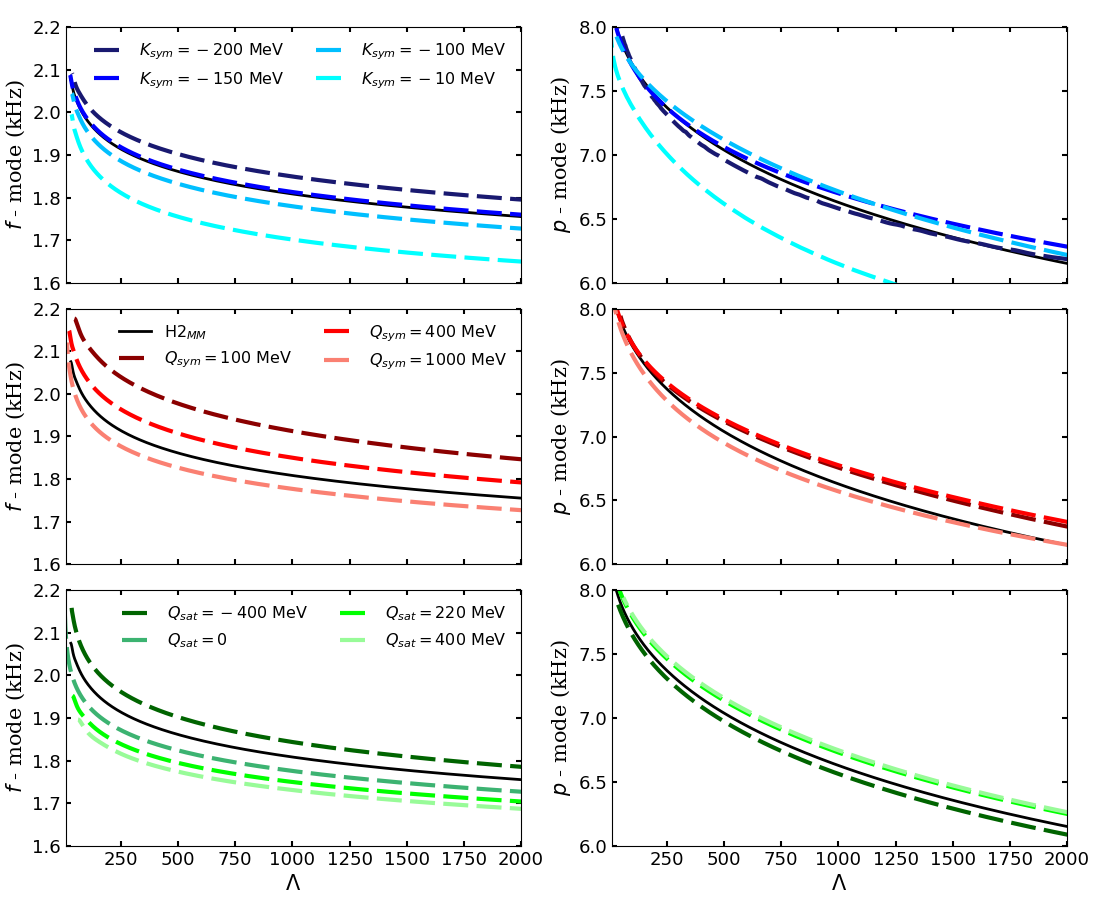} 
\caption{Correlations of the $f$-modes (left panels) and $p$-modes (right panels) with respect to the NS tidal deformability for the EoS used in the present work. The top panels shows the effect of varying $K_\sym$ individually, while the effect of varying $Q_\sym$ is displayed in the center and of varying $Q_\sat$ in the bottom panel. The reference EoS H2$_{MM}$ is shown in all panels with solid black line.}
\label{fig:corrNEP}
\end{figure*}

We analyze in Figure~\ref{fig:corrNEP} the correlations between the NS oscillation modes with the the tidal deformability. The left panels shows that that the $f$-mode has a strong decrease with increasing $\Lambda$, and from values of $\Lambda \approx$ 500 the decrease of the $f$-mode has a slightly decrease with a line behavior.
The $p$-modes (right panels) also decrease with increasing tidal deformability, with a higher variation of $p$-mode with $\Lambda$ in comparison to the $f$-modes.

\section{Conclusions}
\label{sec:conclusion}

In the present work, we have studied the impact of the independent variation of the isovector incompressibility $K_{\rm sym}$, the skewness of the symmetry energy $Q_{\rm sym}$ and the skewness of the binding energy in symmetric nuclear matter $ Q_{\rm sat}$ on the non-radial oscillations of NS.
We have constructed twelve new unified EoS, where the low density controlled by chiral EFT predictions~\cite{Drischler2016} and the crust is modeled within a CLDM~\cite{Grams22a}. 

The impact of the NEP on the $f-$mode oscillations is higher or lower depending on the NS mass: $K_{\rm sym}$ shows an important effect on NS from sub solar masses to 1.3 M$_{\odot}$, $Q_{\rm sat}$ impacts more intermediate NS masses from 1.2 to 1.9 M$_{\odot}$ and $Q_{\rm sym}$ has a strong effect for heavy NS.
In the case of the first $p-$mode, we found a higher impact from $Q_{\rm sat}$ which shows a considerable decrease of this oscillation mode for higher values of  $Q_{\rm sat}$ independent of the NS mass.
For the tidal deformability, the NEP $Q_{\rm sat}$ and $Q_{\rm sym}$ show a bigger impact than $K_{\rm sym}$.
In special, we found that positive values  of $Q_{\rm sat}$ lies outside of the 50$\%$ confidence lines of Ligo-Virgo. It is important to note that these are qualitative results, since we vary independently each parameter while keeping all other NEP fixed to the the reference H$_2$ EoS. It is however a first study that highlight the impact of the high-order NEP on NS asteroseismology.

Given the difficulty to constrain high-order NEP with current nuclear physics experiments and ab-initio theory, the prospect of higher sensitivity of 3G GW detectors shows that the study of GW asteroseiesmology of NS promises an alternative tool to constrain high-order NEP.

\acknowledgments
GG is supported by the Fonds de la Recherche Scientifique (F.R.S.-FNRS) and the Fonds Wetenschappelijk Onderzoek - Vlaanderen (FWO) under the EOS Projects nr O022818F and O000422F. 
CHL would like to thank CNPq for the financial support with Projects No 401565/2023-8 (Universal CNPq) and No 305327/2023-2 (productivity program).
CVF also makes his acknowledgements for the financial support of the productivity program of the
Conselho Nacional de Desenvolvimento Cientıfico e Tecnologico (CNPq), with Project No. 304569/2022-4.

\bibliography{mybib.bib}{}
\end{document}